\documentclass[twocolumn,amsthm]{autart}    
\usepackage{amsfonts}
\usepackage{amsmath, calc}   
\usepackage{graphicx}      
\usepackage{xcolor}   
\usepackage{tikz}
\usepackage{tabularx}
\usepackage{footnote}

\allowdisplaybreaks

\usepackage[ruled,vlined]{algorithm2e}

\begin{document}   
	
\begin{frontmatter}
		
\title{IsoCost-Based Dynamic Programming for Solving Infinite Horizon Optimal Control Problems} 
		
\author[Rahimi]{Saeed Rahimi}\ead{rahimi.saeed@ut.ac.ir}, 
\author[Salimi]{Amir Salimi Lafmejani}\ead{asalimil@asu.edu},    
\author[Kalhor,*,**]{Ahmad Kalhor}\ead{akalhor@ut.ac.ir}

\thanks[**]{The idea of ''IsoCost Surface" with its geometric properties in the optimal control had been initially proposed and used by the corresponding author in the final project of the ''Distributed AI" course at the university of Tehran in 2005.}

\address[Rahimi]{School of Mechanical Engineering, University of Tehran, Tehran, Iran}      
                                 
\address[Salimi]{School of Electrical, Computer and Energy Engineering, Arizona State University, USA} 
        
\address[Kalhor]{Control and Intelligent Processing Center of Excellence, School of Electrical and Computer Engineering, University of Tehran, Tehran, Iran}      

\address[*]{Corresponding Author}

		
\begin{abstract}                          
An innovative numerical algorithm for solving infinite-horizon optimal control problems is introduced in this paper, using the IsoCost-HyperSurface (ICHS) concept. In the state space of an optimal control system, an ICHS is defined as a set of state points having a certain amount of the cost value. In this paper, it is proved that for a certain cost amount, the ICHS resulting from the infinite-horizon optimal control solution, surrounds all other ICHSs resulting from non-optimal control strategies. Regarding this geometric feature, the novel Isocost Dynamic Programming (IDP) algorithm is introduced to search for optimal control solutions. As an illustration of the introduced ICHS concepts and to demonstrate the effectiveness of the proposed IDP algorithm, several simulated examples are presented. The results are compared with those of conventional DP. These comparisons demonstrate that the proposed algorithm has better relative optimality compared to the DP algorithm with 18 \% lower cumulative cost value, by comparing results from the closed-loop control of two nonlinear systems with random initial conditions. More significantly, when compared to the DP algorithm, the IDP was able to enhance computational performance by reducing the execution time by 21 \% while using less memory.
\end{abstract}
	
\end{frontmatter}
	
\section{Introduction}

Developing optimal control strategies for engineering systems relies on solving mathematical problems defined to optimize a long-term scalar cost function. The Bellman's principle of optimality, formulated in the Hamilton-Jacobi-Bellman (HJB) equation, is a principle that various optimization problems has been established on ~\cite{pesch2009maximum}. As well as providing sufficient conditions for optimality, HJB also provides a framework for determining optimal control policies and cost functions~\cite{yang2021hamiltonian}. 
\newline
The literature presents a multitude of methods for solving the HJB optimization problem~\cite{crespo2003stochastic}.
A notable example of these methods, the Linear Quadratic Regulator (LQR), is a benchmark approach in the field of optimal control~\cite{bemporad2002explicit,gupta2018solution,zhang2019inverse}. However, finding closed-form solutions for many optimal control problems is not possible~\cite{sussmann1997300} since finding the analytical solution requires solving complex equations, and even in many cases, there is no guarantee that an analytical solution exists. Consequently, various numerical algorithms have been introduced in recent years to provide feasible solutions to optimal control problems~\cite{rao2009survey}.
\newline
Dynamic Programming (DP) is a family of numerical iterative approaches that exploit the HJB equation~\cite{borrelli2005dynamic} as a tool for evaluating the optimality of solutions.
In the standard DP, a cost value iteration search is used to derive an optimal control policy in the domain of a mesh grid that represents the state space of the system~\cite{ueda2008dynamic}. The versatility of DP methods is attained at the cost of computational complexity and high memory requirements, which are necessary to store optimal policies and value functions for every iteration on the grid~\cite{liu2020adaptive}. Furthermore, as the dimension of the problem increases in size, the computational complexity associated with DP becomes prohibitive, a phenomenon known as the "curse of dimensionality"~\cite{lebedev2022gradient}. 
\newline
Accordingly,~\cite{werbos1992approximate} originally developed Adaptive Dynamic Programming (ADP) algorithms as a way to find an approximate solution of Discrete-Time (DT) HJB~\cite{li2021novel}. 
This method solves the problem of computational complexity, although it suffers from discretization error. More significantly, ADP techniques are not practical for problems with state constraints. This is because the gradient descent method used in the ADP algorithm is only suited for solving unconstrained policy optimization problems~\cite{duan2022adaptive}. In~\cite{mitze2020dynamic}, an algorithm is proposed to find an active optimal set for a constrained LQR control problem with the help of DP. 
\newline
Finding solutions to optimal control problems may be facilitated by utilizing the geometry of the problems~\cite{schattler2012geometric}. 
A differential geometric method for finding explicit closed-form solution for nonlinear optimal control problems using Lie theory is described in~\cite{baillieul1978geometric}. Furthermore,~\cite{kobilarov2011discrete} introduced an approach based on the coordinate-free variational discretization of the dynamics that exploit the nature of the state space to develop a numerical method for finding optimal trajectories for mechanical systems.
\newline
To the best of authors' knowledge, the concept of IsoCost HyperSurface (ICHS) is used for the first time in this paper to develop a novel numerical-geometrical approach for solving infinite horizon optimal control problems. The proposed concept gives an straightforward insight into getting around the impossibility of finding an exact analytical solution and the computational complexities of other numerical algorithms. 
It is proved that for a specific amount of cost, the ICHSs generated by the optimal control solution surround all other ICHSs resulting from non-optimal control solutions. A meta-heuristic local search algorithm is employed in combination with the concept of ICHS to form the Isocost Dynamic Programming (IDP), that aims to search for the optimal control strategies in a reverse iteration manner. 
\newline
The main contribution of IDP approach, as demonstrated through comprehensive examples, is that it can solve the optimal control problem for a broad range of dynamical systems, from linear to nonlinear systems. 
The second major contribution is that this method reduces the computational complexity of solving optimization problems, compared to other numerical methods. As a third contribution, the proposed IDP controller improves the performance and accuracy of the system by increasing the relative optimality of the solutions by reducing terminal cost value. An assessment of the contribution mentioned is made by comparing IDP method with standard DP for two different nonlinear systems.
\newline
In this paper, Section.~\ref{sec statement} introduces the problem formulations and definitions required for establishing the ICHS concept. The main results of the IsoCost Based DP method, based on earlier definitions, are then provided in Section.~\ref{sec main results}. Furthermore, a framework of algorithms for achieving ICHS and designing IDP controller is suggested in this section. In order to evaluate the performance of the IDP method in the closed-loop control of dynamical systems, two different optimal control problems are simulated in Section.~\ref{sec simul}. Additionally, the results from IDP controller is also compared with conventional DP and LQR controllers. Finally, in Section.~\ref{sec concl}, the conclusion and insights are given.

\begin{figure*}[t]
	\begin{center}
		
		\includegraphics[scale=0.45]{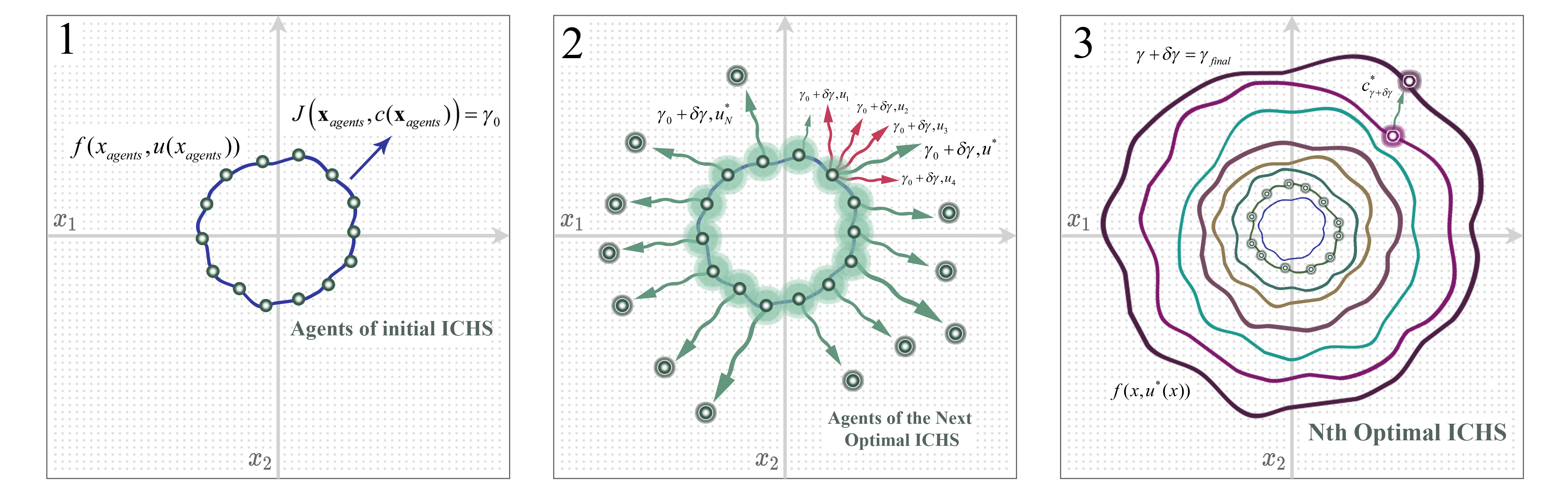}    
\caption{The illustration of IDP searching for optimal control policies in the two dimensional state space of a system. A large number of state points are chosen as initial agents. These agents can be chosen randomly near the origin. For each initial agent, the optimal policy is the one that drives the initial agent further from the origin, resulting in the maximization of the ICHS area formed by all agents. The IDP algorithm continues to explore the state space of the system, and storing optimal policies.} 
		\label{fig1} 
	\end{center}                               
\end{figure*}
\section{Problem Statement and Preliminaries}\label{sec statement}
This section defines the optimal control problem and ICHS formulations for a dynamical system. These definitions lay the foundation for incorporating ICHS into optimal control design.
\newline
Consider the following differential equation of a deterministic nonlinear system:	
\begin{equation}\label{eq1}
		\dot{\boldsymbol{x}}(t)=\boldsymbol{f}(\boldsymbol{x}(t), \boldsymbol{u}(t)), \quad \boldsymbol{f}(0,0)=0
\end{equation}
where $\boldsymbol{x} \in \mathbb{R}^{n}, \boldsymbol{u} \in \mathbb{R}^{m}$ and $\boldsymbol{f}: \mathbb{R}^{n} \times \mathbb{R}^{m} \rightarrow\mathbb{R}^{n}$ ‌denote the state vector, input control and a vector of continuously differentiable functions, respectively. The following assumptions are made for a system represented in Eq.~(\ref{eq1}).
	
\textbf{Assumption 1.}
The system in Eq.~(\ref{eq1}) is deterministic, unconstrained and controllable in a certain continuous region $D_{R} \subset \mathbb{R}^{n}$ including the origin. Moreover, for the aforementioned nonlinear system, infinite horizon cost function is defined as:
\begin{equation}\label{eq2}
		J(\boldsymbol{x}(t), \boldsymbol{u})=\int_{0}^{\infty} g(\boldsymbol{x}(\tau), \boldsymbol{u}(\tau)) d \tau~, \quad g(0,0)=0
\end{equation}
where $g(.,.)$ is a smooth positive definite function.

\textbf{Assumption 2.}
The control solution $\boldsymbol{u} = c(\boldsymbol{x})$ asymptotically stabilizes the system in Eq.~(\ref{eq1}) to the origin with the domain of attraction, $D_{c} \subset D_{R}$.

\textbf{Definition 1.}
Given the state space system in Eq.~(\ref{eq1}) with the cost function in Eq.~(\ref{eq2}) and under Assumptions~1 and 2, for all positive real scalar values $\gamma$, an ICHS is defined as follows:
\begin{equation}\label{eq3}
		S_{\gamma}(\boldsymbol{x}, c(\boldsymbol{x}))=\left\{\boldsymbol{x} \mid \boldsymbol{x} \in D_{c}~~, \quad J(\boldsymbol{x}(t), c(\boldsymbol{x}))=\gamma\right\}
\end{equation}

\textbf{Definition 2.}
For the system described in Eq.~(\ref{eq1}) and Eq.~(\ref{eq2}) and Assumptions~1 and~2, an IsoCost HyperVolume (ICHV) is defined as:
\begin{equation}\label{eq4}
		\Omega_{\gamma}(\boldsymbol{x}, c(\boldsymbol{x}))=\left\{\boldsymbol{x} \mid \boldsymbol{x} \in D_{c}~~, \quad J(\boldsymbol{x}(t), c(\boldsymbol{x})) < \gamma\right\}
\end{equation}
The ICHV gives a criterion to examine the relative optimality of different control solutions in state space. 

\textbf{Remark 1.}
 For the system defined in Eq.~(\ref{eq1}) with the cost function according to Eq.~(\ref{eq2}) and regarding to Definition~2 and aforementioned assumptions, it is understood that for two different control strategies $c_{i}(\boldsymbol{x})$ and $c_{j}(\boldsymbol{x})$ and each positive real scalar $\gamma$ it has:
\begin{equation}
			\Psi_{\gamma}^{i}\left(\boldsymbol{x}, {c_{i}(\boldsymbol{x})}, {c_{j}(\boldsymbol{x})}\right)= 
			\Omega_{\gamma}\left(\boldsymbol{x}, c_{i}(\boldsymbol{x})\right)-\Omega_{\gamma}\left(\boldsymbol{x}, c_{j}(\boldsymbol{x})\right) 
\end{equation}
determine sub spaces of the state space, in which $c_{i}(\boldsymbol{x})$ made less cost function value compared to $c_{j}(\boldsymbol{x})$, in other words, the control $c_{i}(\boldsymbol{x})$ corresponds to a larger HyperVolume (HV) compared to that of corresponding to $c_{j}(\boldsymbol{x})$. Respectively, $\Psi_{\gamma}^{j}$ is defined as:
\begin{equation}
	\Psi_{\gamma}^{j}\left(\boldsymbol{x}, {c_{j}(\boldsymbol{x})}, {c_{i}(\boldsymbol{x})}\right)= 
	\Omega_{\gamma}\left(\boldsymbol{x}, c_{j}(\boldsymbol{x})\right)-\Omega_{\gamma}\left(\boldsymbol{x}, c_{i}(\boldsymbol{x})\right) 
\end{equation}
corresponding to sub spaces of state space in which $c_{j}(\boldsymbol{x})$ makes a bigger HV. Subsequently:
\begin{equation}
	\begin{aligned}
		\forall \boldsymbol{x}(t) \in \Psi_{\gamma}^{i}\left(\boldsymbol{x}, c_{i}(\boldsymbol{x}), c_{j}(\boldsymbol{x})\right) & \Rightarrow \\
		J\left(\boldsymbol{x}(t), c_{i}(\boldsymbol{x})\right) &<J\left(\boldsymbol{x}(t), c_{j}(\boldsymbol{x})\right) \\
		\forall \boldsymbol{x}(t) \in \Psi_{\gamma}^{j}\left(\boldsymbol{x}, c_{j}(\boldsymbol{x}), c_{i}(\boldsymbol{x})\right) & \Rightarrow \\
		J\left(\boldsymbol{x}(t), c_{j}(\boldsymbol{x})\right) &<J\left(\boldsymbol{x}(t), c_{i}(\boldsymbol{x})\right)
	\end{aligned}
\end{equation}
\textbf{Definition 3.}
Consider there are two different control solutions $c_i(\boldsymbol{x})$ and $c_j(\boldsymbol{x})$ satisfying the assumptions for the system in Eqs.~(\ref{eq1}) and~(\ref{eq2}). For a positive real scalar $\gamma$, through which $S_{\gamma}\left(\boldsymbol{x}, c_{i}(\boldsymbol{x})\right)$ is not a null set, $S_{\gamma}\left(\boldsymbol{x}, c_{i}(\boldsymbol{x})\right)$ surrounds $S_{\gamma}\left(\boldsymbol{x}, c_{j}(\boldsymbol{x})\right)$ if:
\begin{equation}\label{eq8}
		S_{\gamma}\left(\boldsymbol{x}, c_{i}(\boldsymbol{x})\right) \cap \Omega_{\gamma}\left(\boldsymbol{x}, c_{j}(\boldsymbol{x})\right)=\emptyset
\end{equation}
\textbf{Remark 2.}
With regard to Eqs.~\eqref{eq4} and~\eqref{eq8}, for each control strategy, the corresponding ICHS surrounds itself. 

\textbf{Definition 4.} 
Two Control strategies $c_{i}(\boldsymbol{x})$ and $c_{j}(\boldsymbol{x})$ are equal for the system in Eq.~\eqref{eq1} and the cost function Eq.~\eqref{eq2} if their ICHSs are equal for each $\gamma$.

With the above assumptions and definitions in hand, the ICHSs can be achieved for any nonlinear system with a specified cost function. 

\section{Main Results}\label{sec main results}
Solving optimal control problems based on the theory of optimal ICHS is the focus of this section. This section begins by discussing the fundamental theorem of this research as the main results. Then, the theory of ICHS will be put into practice, by introducing the IDP controller. 

\textbf{Theorem 1 (Theory of ICHS Optimality).}
\textit{Consider $c^{*}(\boldsymbol{x})$ is the optimal control solution of the system with Eq.~(\ref{eq1}), minimizing the cost function according to Eq.~(\ref{eq2}). For each positive real scalar $\gamma$, $S_{\gamma}\left(\boldsymbol{x}, c^{*}(\boldsymbol{x})\right)$ surrounds any other IsoCost HyperSurfaces $S_{\gamma}\left(\boldsymbol{x}, c(\boldsymbol{x})\right)$ that are obtained by non-optimal control solutions $c(\boldsymbol{x})$.}
	
\textbf{Proof.} If $S_{\gamma}\left(\boldsymbol{x}, c^{*}(\boldsymbol{x})\right)$ does not surround $S_{\gamma}\left(\boldsymbol{x}, c(\boldsymbol{x})\right)$, regarding to Definition~3, there is at least one state such as $\boldsymbol{x}_{0}$ which simultaneously belongs to both $S_{\gamma}\left(\boldsymbol{x}, c^{*}(\boldsymbol{x})\right)$ and $\Omega_{\gamma}\left(\boldsymbol{x}, c(\boldsymbol{x})\right)$. According to Definition~1 and for the optimal control solution $c^{*}(\boldsymbol{x})$, since $\boldsymbol{x}_{0} \in S_{\gamma}\left(\boldsymbol{x}, c^{*}(\boldsymbol{x})\right)$ then:
\begin{equation}\label{eq9}
		J\left(\boldsymbol{x}_{0}, c^{*}(\boldsymbol{x})\right)=\gamma
\end{equation}

By considering the Definition~2, for the non-optimal control solutions $c(\boldsymbol{x})$, since $\boldsymbol{x}_{0} \in \Omega_{\gamma}(\boldsymbol{x}, c(\boldsymbol{x}))$ then:	
\begin{equation}\label{eq10}
	 \boldsymbol{x}_{0} \in \Omega_{\gamma}(\boldsymbol{x}, c(\boldsymbol{x})) \quad \Rightarrow \quad J\left(\boldsymbol{x}_{0}, c(\boldsymbol{x})\right) < \gamma
\end{equation}
It is concluded that for the state $\boldsymbol{x}_{0}$, the non-optimal control solution $c(\boldsymbol{x})$ provides the less amount of cost, in comparison with the optimal control $c^{*}(\boldsymbol{x})$. Therefore, this is against the considered assumption about the optimality of $c^{*}(\boldsymbol{x})$. Consequently, intersection of $S_{\gamma}\left(\boldsymbol{x}, c^{*}(\boldsymbol{x})\right)$ and $\Omega_{\gamma}(\boldsymbol{x}, c(\boldsymbol{x}))$ must be a null set and regarding to Definition~3, $S_{\gamma}\left(\boldsymbol{x}, c^{*}(\boldsymbol{x})\right)$ surrounds $S_{\gamma}\left(\boldsymbol{x}, c(\boldsymbol{x})\right)$. \qedsymbol
	
\textbf{Remark 3} Regarding to Definition~1 to~3 and Remark~1, an important result of Theorem.~1 is that for each positive real scalar $\gamma$, the ICHV $\Omega_{\gamma}\left(\boldsymbol{x}, c^{*}(\boldsymbol{x})\right)$ includes all CHVs of the non-optimal control solutions:
\begin{equation}
	\forall c(\boldsymbol{x}) \notin c^{*}(\boldsymbol{x}): \quad \Omega_{\gamma}(\boldsymbol{x}, c(\boldsymbol{x})) \subset \Omega_{\gamma}\left(\boldsymbol{x}, c^{*}(\boldsymbol{x})\right)
\end{equation}
Therefore, one can say that for each positive real scalar $\gamma$ the ICHV corresponding to the optimal control solution has been maximized in comparison to other ICHVs of non-optimal control solutions.

\textbf{Definition 4.} For the optimal control problem stated in Eqs.~(\ref{eq1}) and~(\ref{eq2}) and for all state points in $V_{\gamma}\left(\boldsymbol{x}, c^{*}(\boldsymbol{x})\right)$, the control solution $c_{\gamma}^{*}(\boldsymbol{x})$ is the same optimal control solution $c^{*}(\boldsymbol{x})$.

\textbf{Remark 4.}
Regarding to Remark~3 and for a certain amount of $\gamma$, the optimal control solution $c_{\gamma}^{*}(\mathbf{x})$ must be explored in order that maximum possible ICHV is achieved:
\begin{equation}\label{eq11}
	c_{\gamma}^{*}=\arg _{c(\boldsymbol{x})}\left(\max \left(V_{\gamma}(\boldsymbol{x}, c(\boldsymbol{x}))\right)\right)
\end{equation}
Using the definition of optimal ICHS, as well as Definition~2 and Remark~4, the optimal control problem is reduced to a problem aimed at maximizing the volume $V_{\gamma}\left(\boldsymbol{x}, c^{*}(\boldsymbol{x})\right)$ of a control solution.
\newline
However, exploring the optimal control that can satisfy Eq.~(\ref{eq11}) is intractable since for each control strategy, it is required to estimate $V_{\gamma}(\boldsymbol{x}, c(\boldsymbol{x}))$, which is not computationally efficient for high dimension problems. 
As the nonlinear system Eq.~(\ref{eq1}) and the cost function Eq.~(\ref{eq2}) are assumed to be continuous, the optimal control solution is also continuous. Hence, using the concept of ICHSs, it is possible to design a recursive version of Eq.~(\ref{eq11}), resulting in a significant reduction of computational load.
	
\textbf{Remark 5.} 
For a certain amount of $\gamma$ and a small enough $d \gamma$, assume the optimal control solution $c_{\gamma}^{*}(\boldsymbol{x})$ and the ICHS, $S_{\gamma}\left(\boldsymbol{x}, c_{\gamma}^{*}(\boldsymbol{x})\right)$ are known. To estimate $S_{\gamma+d \gamma}\left(\boldsymbol{x}, c_{\gamma+d \gamma}^{*}(\boldsymbol{x})\right)$, the optimal control solution $c_{\gamma+d \gamma}^{*}(\boldsymbol{x})$ can be explored in a restricted interval around $c_{\gamma}^{*}(\boldsymbol{x})$ via maximizing $V_{\gamma+d \gamma}(\boldsymbol{x}, c(\boldsymbol{x}))$.

Knowing an initial set of optimal ICHSs and taking Remark~5 into account, the state space of the system can be explored for optimal control solutions.
Although the aforementioned idea appears to be feasible, some critical aspects must be addressed. How the initial optimal ICHS, $S_{\gamma_{0}}\left(x, c_{\gamma_{0}}^{*}(\boldsymbol{x})\right)$ should be found and which exploring algorithms can be used to search state space of the systems and compute ICHVs?
\newline
The Determination of the first ICHS as an initial solution is an essential criterion in the proposed approach. The optimal solutions could not be found correctly if the initial solution would not be appropriately determined.
As a result, by considering a small enough boundary about origin for a small amount of cost $\gamma_0$, the agents can be selected randomly to shape the first ICHS, $S_{\gamma_{0}}\left(\boldsymbol{x}, c_{\gamma_{0}}^{*}(\boldsymbol{x})\right)$ for the problem. 
\newline
However, the initial ICHS can be determined using the LQR technique if Eq.~(\ref{eq1}) can be linearized about the origin and the cost function Eq.~(\ref{eq2}) can be approximated in a quadratic form.
Then, by discretizing the optimal control problem and using a search method like Genetics Algorithm (GA) in a reverse trajectory manner, the next optimal ICHS can be found by maximizing the area of the surface made by the points on the boundary of the initial ICHS referring to Theorem~1. Furthermore, agents located on the IsoCost surface boundary move to the boundary of the next IsoCost surface to explore the remaining state space of the system for optimal solutions.
\SetKwComment{Comment}{}{~$\triangleleft $}
\SetKwComment{Commenttt}{$\triangleright $}{}
\SetKwInput{kwInit}{Initialize}
\SetKwInput{kwout}{Output}
\SetKwInput{kwDef}{Define}
\SetKwInput{kwCons}{Consider}
\SetKwInput{kwDt}{2:~Data}
\SetKwInput{kwInit}{Initialization}
\SetKwInput{kwSet}{1:~Set}
\begin{algorithm}[h]\label{alg1}
	\DontPrintSemicolon
	\newcommand\mycommfont[1]{\footnotesize\ttfamily\textcolor{cyan}{#1}}
	\SetCommentSty{mycommfont}
	\caption{Estimation of ICHS $S_{\gamma_{k+1}}\left(\boldsymbol{x}, c(\boldsymbol{x})\right)$ for a system described in Eqs.~(\ref{eq1}) and~(\ref{eq2}) }
	\vspace{0.2em}
    \vspace{0.2em}
    \textbf{1:}~\kwInit{\Comment*[r]{Define the starting ICHS}}{
    \vspace{0.1em}
    \eIf{k=0 
    \vspace{0.1em}}
    {
	{\textbf{Define}: {$N$\Comment*[r]{number of agents on the initial ICHS}
					   \Indp \Indp \Indp ~ $R_{0}$~~\Comment*[r]{radius of initial random surface} 
	 				   ~~~$\gamma_{0}$~~\Comment*[r]{initial ICHS cost value}}}
 	\vspace{0.1em}		   
    $X_{R_{0}} \leftarrow$ \text{N random points on a circle with radius $R_0$} \;
    \vspace{0.1em}	
    $S_{\gamma_{0}}\left(\boldsymbol{x}, c(\boldsymbol{x})\right)\leftarrow X_{R_{0}}$ 
    \vspace{0.1em}		      
    }
    {\Comment*[r]{Use previous ICHS as starting surface}
    \vspace{0.2em}	
    $X_{R_{k}} \leftarrow S_{\gamma_{k}}\left(\boldsymbol{x}, c(\boldsymbol{x})\right) , ~(k=1, 2, 3, 4,...)$}
    \vspace{0.2em}
	}
	\vspace{0.2em}
    \textbf{2:}~\kwDef{${u} \leftarrow c$ \Comment*[r]{define a control strategy}}
	\vspace{0.2em}
	\textbf{3:}~\For{$\boldsymbol{x} \in \boldsymbol{X}_{R_{k}}$}{
	\vspace{0.2em}
	~$\boldsymbol{x}_{T} \leftarrow \boldsymbol{x} $ \;
	\vspace{0.2em}
	~$R_{k+1} \leftarrow \frac{1}{N} \Sigma~\|\boldsymbol{X}_{R_{k}}\|^2 $ \;	
	\vspace{0.2em}	
	~$\gamma_{k+1} \leftarrow (\frac{R_{k+1}}{R_{k}})^2  \gamma_{k} $ \Comment*[r]{cost value for surface $R_{k}$}
	\vspace{0.2em}
	~$g_\text{c} \leftarrow g(\boldsymbol{x}_{T}, c(\boldsymbol{x}_{T}))$	\Comment*[r]{g: cost function}
	\vspace{0.2em}
	~$\Delta t \leftarrow -\frac{\gamma_{k+1}}{g_\text{c}}$\Comment*[r]{time step for backward iteration}
	\vspace{0.2em}
	~$\dot{\boldsymbol{x}}_{T}\leftarrow\boldsymbol{f}(\boldsymbol{x}_{T}, {c}(\boldsymbol{x}_{T}))$\;
	\vspace{0.2em}
	~$ \boldsymbol{x}_{T-\Delta t}$ $\leftarrow$ \text{Solve Eq.~(\ref{eq1}) for (${\Delta t},~\boldsymbol{x}_{T},~ \dot{\boldsymbol{x}}_{T},~c $)}\;
	\vspace{0.2em}
	~$\boldsymbol{x}_{\gamma_{k+1}} \leftarrow \boldsymbol{{x}}_{T-\Delta t}$\Comment*[r]{achieved next isocost state}
	\vspace{0.2em}
	~$\boldsymbol{X}_{\gamma_{k+1}} \leftarrow \boldsymbol{x}_{\gamma_{k+1}}$\Comment*[r]{store all next isocost states}
	}
	\vspace{0.2em}
	\textbf{4:}~\kwout{$S_{\gamma_{k+1}}\left(\boldsymbol{x}, c(\boldsymbol{x})\right) \leftarrow \boldsymbol{X}_{\gamma_{k+1}}$\Comment*[r]{next ICHS}} 
\end{algorithm}
\newline
Algorithms~\ref{alg1} and~\ref{alg2} are provided to outline the main results presented in theorem and remarks. Moreover, Fig.~\ref{fig1} is an illustration of the main results of the paper and mentioned algorithms. Algorithm~\ref{alg1} calculates the ICHS for any cost value $\gamma$ and control solution $c(\boldsymbol{x})$. In addition, the Algorithm~\ref{alg2} provides a framework for designing an optimal IDP controller for the system described in Eqs.~(\ref{eq1}) and~(\ref{eq2}). By considering an interval for control effort, search methods like GA can explore controller inputs that drive each state further from the initial state, resulting in the maximization of next ICHS area. 
As a result, Algorithm~2 gives a frame work for calculating optimal control $c^{*}$ for each state using the concept of IDP. The derived values of $c^{*}$ optimized by GA, for each state, can be saved into the memory and used as a closed-loop feedback control law.

\RestyleAlgo{ruled}
\SetKwComment{Comment}{}{~$\triangleleft $}
\SetKwComment{Commenttt}{$\triangleright $}{}
\SetKwInput{kwInit}{Initialization}
\SetKwInput{kwout}{Output}
\SetKwInput{kwDef}{Define}
\SetKwInput{kwCons}{Consider}
\SetKwInput{kwSet}{Set}
\SetKwInput{kwIn}{Inputs}
\SetKwInput{kwreturn}{Return}
\begin{algorithm}[h]\label{alg2}
	\DontPrintSemicolon
	\newcommand\mycommfont[1]{\footnotesize\ttfamily\textcolor{cyan}{#1}}
	\SetCommentSty{mycommfont}
	\caption{Finding optimal control solution for Eqs.~(\ref{eq1}) and~(\ref{eq2}) using IsoCost Dynamic Programming}
	\vspace{0.2em}
	\textbf{1:}~\kwSet{~${\gamma_{f},~\gamma_{0},~N,~R_0}$\Comment*[r]{$\gamma_{f}: \text{terminal cost value}$}}
	\vspace{0.25em}
	\textbf{2:}~\kwInit{\Comment*[r]{obtaining initial ICHS}}{
	\vspace{0.25em}
	$\gamma_{c} \leftarrow \gamma_{0}~~~(\text{From Algorithm~\ref{alg1}})$\Comment*[r]{$\gamma_{c}: \text{current cost value}$}
	\vspace{0.2em}
	 $S_{\gamma_0}\left(\boldsymbol{x}, c^{*}(\boldsymbol{x})\right) \leftarrow S_{\gamma_0}\left(\boldsymbol{x}, c(\boldsymbol{x})\right)~~~(\text{From Algorithm~\ref{alg1}})$
	} \;
	\vspace{0.25em}	
	\textbf{3:}~\kwDef{$\boldsymbol{u}$\Comment*[r]{candidate control interval/strategy}
	{
	\vspace{0.25em}
	$\boldsymbol{u} \leftarrow [~u_\text{min}~,~u_\text{max}~]$\Comment*[r]{define an interval for U}
	}}
	\vspace{0.25em}
	k $\leftarrow$ 0 \;
	\vspace{0.25em}
	\While{$\gamma_{c} < \gamma_{f}$}{
	\For{$\boldsymbol{x}~\in~S_{\gamma_{k}}\left(\boldsymbol{x}, c^{*}(\boldsymbol{x})\right)$}{
	\For{${c} \in \boldsymbol{u}$}{
	\vspace{0.25em}
	$\boldsymbol{x}_{\gamma_{k+1}} \leftarrow \text{Step 3 of Algorithm~\ref{alg1}}$ 
	\vspace{0.25em}
	$c_{\gamma_{k+1}}^{*}(x)\leftarrow \arg _{c(\boldsymbol{x})}\left(\max\|\boldsymbol{x}_{\gamma_{k+1}}\|  )\right)$
	
	$\boldsymbol{x}_{\gamma_{k+1}}^{*} \leftarrow \arg _{\boldsymbol{x}}\left(\max\|\boldsymbol{x}_{\gamma_{k+1}}\|)\right)$
	
	Save $c_{\gamma_{k+1}}^{*}(x)$ to $c_{\gamma_{k+1}}^{*}$
	
	Save $\boldsymbol{x}_{\gamma_{k+1}}^{*}$ to $S_{\gamma_{k+1}}\left(\boldsymbol{x}, c^{*}(\boldsymbol{x})\right)$		
	
	}
	\vspace{0.3em}
	}
	Save $S_{\gamma_{k+1}}\left(\boldsymbol{x}, c^{*}(\boldsymbol{x})\right)$ and $c_{\gamma_{k+1}}^{*}(x)$ \;
	k $\leftarrow$ k+1~~,~~$\gamma_{c} \leftarrow \gamma_{k+1}$ 
	}

	\vspace{0.2em}
\end{algorithm}
\begin{figure}[th]
	\begin{center}
		\includegraphics[scale=0.5]{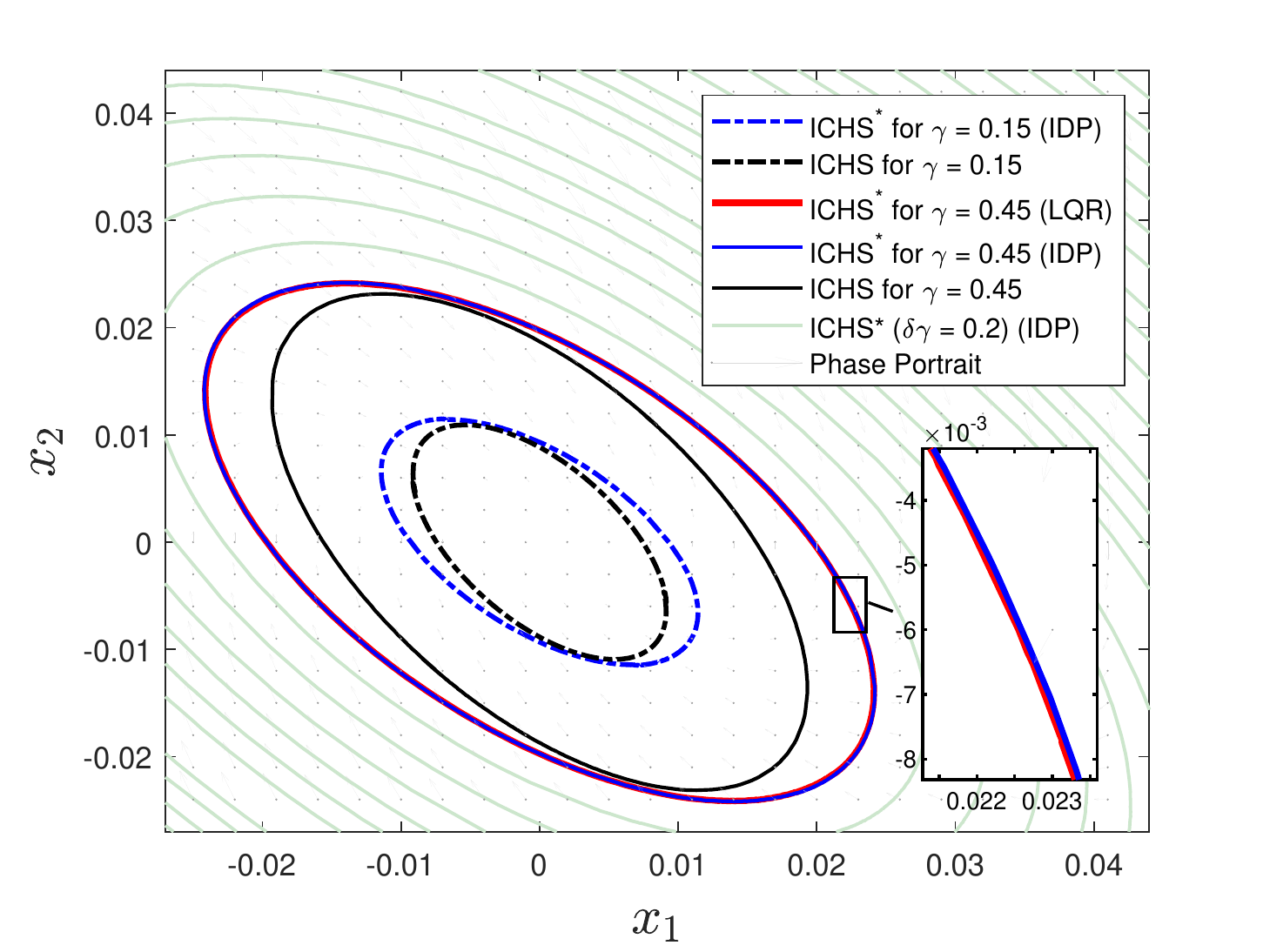}    
		\caption{Resulted ICHSs for system in Eqs.~(\ref{eq12}) and~(\ref{eq13}) using IDP method, for different cost values and control strategies.}  
		\label{fig2}
	\end{center}                                
\end{figure}
\section{Simulations and Comparison Results}\label{sec simul}
In this section, two simulation examples are used to show how the IDP approach can be used to solve an optimal control problem. The goal of the first example is to identify the optimal control strategy for a second order nonlinear system with a nonlinear cost function. The second example, on the other hand, resolves the optimal control problem for a nonlinear inverted pendulum system. The IDP controller and DP technique are compared in both examples. Additionally, a comparison with approximated closed-form solutions is given. For the system that can be linearized and its cost function can be approximated with a quadratic form, initially the system is linearized by applying appropriate diffeomorphism, that converts the nonlinear problem into an optimal quadratic form. Then, the transformed LQR problem is solved and the results are compared with IDP controller.
\newline
First, consider the following second order nonlinear system:
\begin{equation}\label{eq12}
	\begin{aligned}
		&\dot{x}_{1}=x_{2} \\
		&\dot{x}_{2}=u-x_{1}^{2}
	\end{aligned}
\end{equation}

and nonlinear cost function:
\begin{equation}\label{eq13}
	\begin{aligned}
		&J\left({x_{1}}, {x_{2}}, {u}\right)=\int_{0}^{\infty}\left[x_{1}^{2}(\tau)+\sin ^{2}\left(x_{2}(\tau)\right)\right. \\
		&+\cos ^{2}\left(x_{2}(\tau)\right)\left(u(\tau)-x_{1}^{2}(\tau)\right)^{2}] d \tau
	\end{aligned}
\end{equation}
In Fig.~\ref{fig2}, the optimal and non-optimal ICHS for the system in Eqs.~(\ref{eq12}) and~(\ref{eq13}) are represented. As it is expected from Theorem.~1, the optimal ICHS ($ICHS^*$) contours $S_{\gamma}\left(\boldsymbol{x}, c^{*}(\boldsymbol{x})\right)$ from IDP controller  surround the non-optimal ICHSs resulted from non-optimal controllers. Moreover, to show the optimality of ICHSs resulted from IDP, the optimal ICHS from LQR controller are also depicted.  To achieve analytic solution for this example, by using diffeomorphism for Eqs.~(\ref{eq12}) and~(\ref{eq13}), one has:

\begin{equation}\label{eq16}
	\begin{aligned}
		&z_{1} =x_{1} \\
		&z_{2} =\sin \left(x_{2}\right) \\
		&v =\cos \left(x_{2}\right)\left(u-x_{1}^{2}\right)
	\end{aligned}
\end{equation}

So, the nonlinear optimal control problem will transform to an LQR one as follows:
\begin{equation}
	\begin{gathered}
		\dot{\boldsymbol{z}}=\left[\begin{array}{ll}
			0 & 1 \\
			0 & 0
		\end{array}\right] \boldsymbol{z}+\left[\begin{array}{l}
			0 \\
			1
		\end{array}\right] v \\
		g(\boldsymbol{z}, v)=\boldsymbol{z}^{T} \boldsymbol{z}+v^{2}
	\end{gathered}
\end{equation}
\begin{figure}[t]
	\begin{center}
		\includegraphics[scale=0.5]{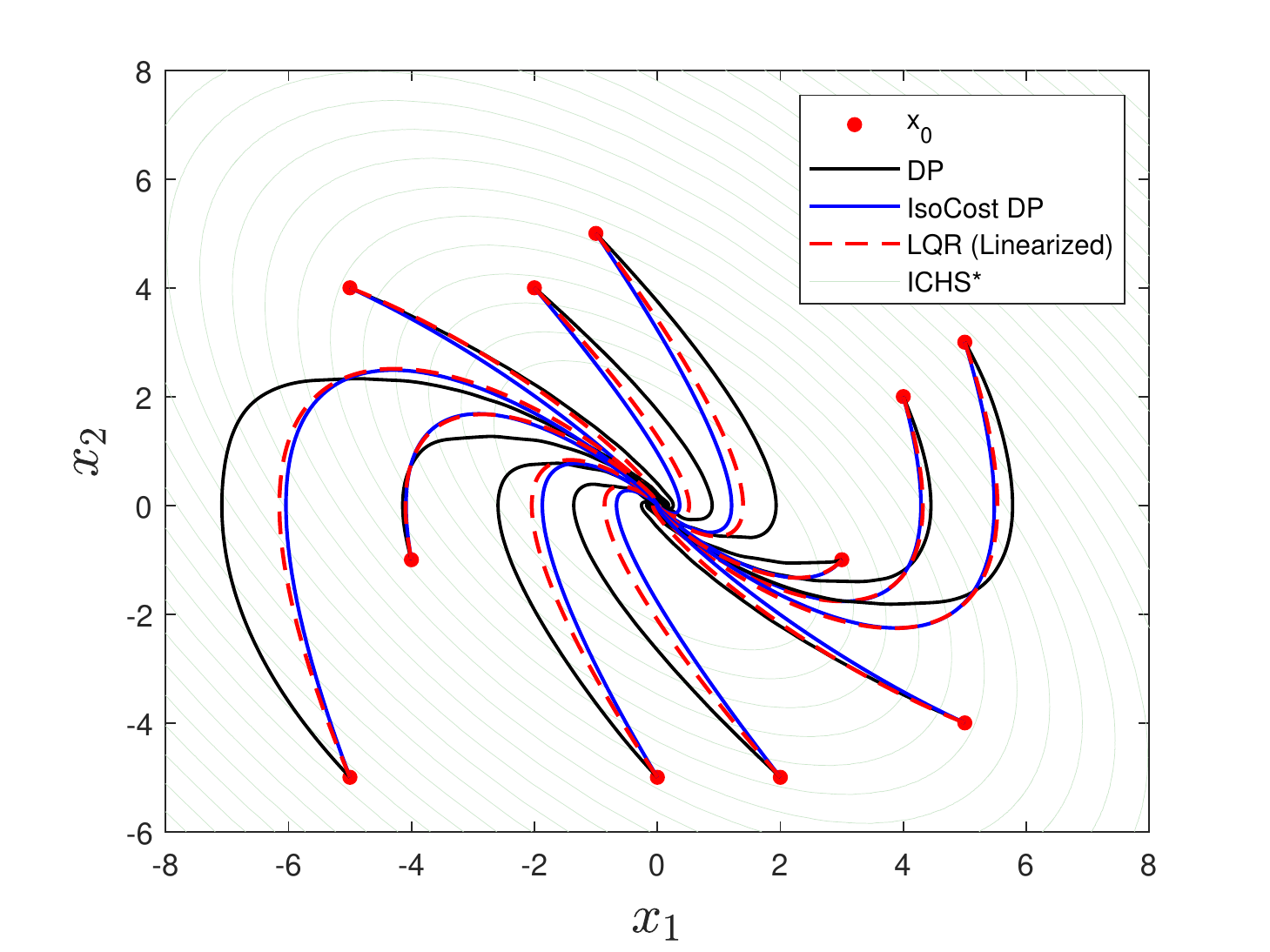}    
		\caption{Comparison between different control strategies in stabilizing the system in Eqs.~(\ref{eq12}) and~(\ref{eq13}) for randomly chosen initial conditions.} 
		\label{fig3} 
	\end{center}                               
\end{figure}
By solving the transformed LQR problem, using Riccati equation, the optimal solution for the LQR problem is computed as $c^{*}(\boldsymbol{z})=-\left[\begin{array}{ll}1 & 1.732\end{array}\right] \boldsymbol{z}$ and then by using inverse of diffeomorphism for Eq.~(\ref{eq16}), the optimal solution $c^{*}(\boldsymbol{x})=\cos ^{-1}\left(x_{2}\right)\left(-x_{1}-1.732 \sin \left(x_{2}\right)+x_{1}^{2}\right)$ is computed. Then, the optimal ICHS is determined for cost value $\gamma$=0.45 which is shown in Fig~\ref{fig2}.
As demonstrated in Fig.~\ref{fig2}, by comparing the $ICHS^*$ from IDP and LQR, it can be concluded that the IDP estimated the optimal ICHS with high accuracy. Additionally, the other $ICHS^*$ according to different cost values are also depicted by considering an increment for cost value ($\delta \gamma$ = 0.2), to show how $ICHS^*$ can cover up the state space of the system and explore the optimal IDP controller for each desired state.  
\newline
Considering Eqs.~(\ref{eq12}) and~(\ref{eq13}), the optimal control solution is found using Algorithms~\ref{alg1} and~\ref{alg2}. The resulted optimal IDP controller is evaluated for 11 randomly chosen initial conditions ($x_0$) as depicted in Fig.~\ref{fig3}. The DP is also used to control the aforementioned system and the results are depicted in Fig.~\ref{fig3} in a comparison with IDP and LQR control. With better accuracy than the DP control, the IDP controller was able to stabilize all initial condition point. This figure also shows the optimal ICHS contours with semi-transparent green color, achieved using Algorithms~1 and~2. 
\newline
\begin{figure}[t]
	\begin{center}
		\includegraphics[scale=0.5]{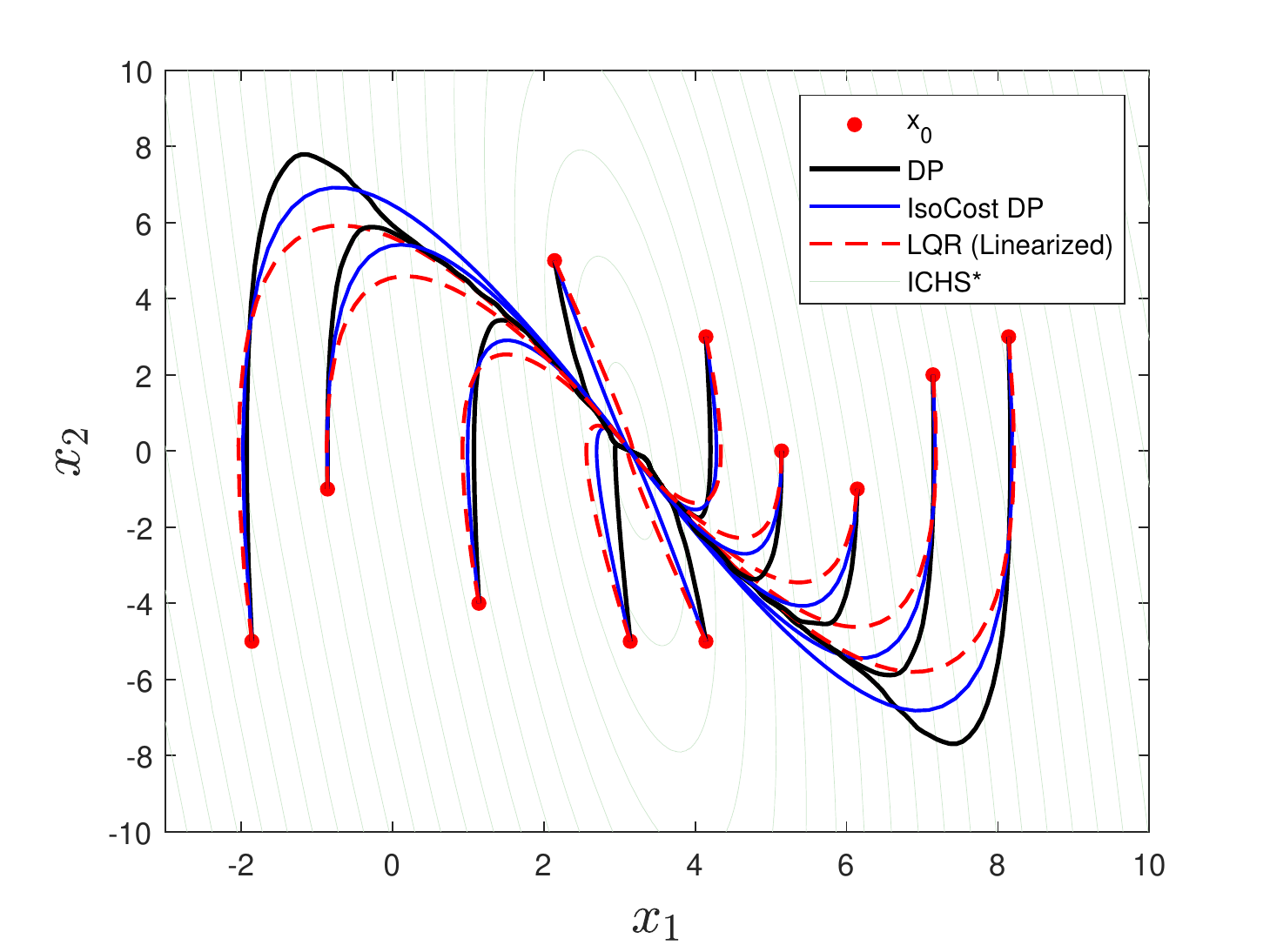}     
		\caption{Comparison between different control strategies in stabilizing the inverted pendulum system for randomly chosen initial conditions.} 
		\label{fig4}
	\end{center}                               
\end{figure}
For the second example, the inverted pendulum problem is utilized. The highly nonlinear dynamic nature of inverted pendulum makes this system a classical benchmark for evaluating different control techniques. The dynamic model of the system can be written as:
\begin{equation}\label{eq14}
	\begin{aligned}
		& x_1 = \theta~,~~~x_2 = \dot{\theta} \\
		&\dot{x}_{1}=x_{2} \\
		&\dot{x}_{2}=\frac{g}{l}sin(x_1) -\frac{b}{ml^2}x_2-\frac{1}{ml^2}u
	\end{aligned}
\end{equation}
where the parameters are considered as $m = 1~kg$, $l = 1~m$ , $b=0.5$ (friction damping coefficient), and $g = 9.81~m/s^2$. The goal is to stabilize the inverted pendulum in a upward standing position with zero velocity ($x_1= \pi~,~~x_2=0 $). The cost function for this optimal control problem is a in the following quadratic form:
\begin{equation}\label{eq15}
	J\left({x_{1}}, {x_{2}}, {u}\right)=\int_{0}^\infty\left(x_{1}^{2}(\tau)+x_{2}^{2}(\tau)+u^2(\tau)\right) d \tau
\end{equation}
\newline
Considering the approach mentioned in Algorithm~1, the optimal ICHSs for system described in Eqs.~(\ref{eq14}) and~(\ref{eq15}) can be achieved using the IDP. The resulted contours are portrayed in Fig.~\ref{fig4} with semi-transparent green color. Moreover, the controlled response of the system to 11 randomly chosen initial conditions is also depicted in Fig.~\ref{fig4}. 
In the simulation of inverted pendulum, the parameters stated in Table~\ref{table1-dp param} are considered for DP and IDP.
This table also shows the parameters used in GA and Algorithm~2 to find the optimal control policy for inverted pendulum system.
As presented, the IDP has more accuracy in controlling system compared to DP. By comparing the behavior of cost function value for one of this initial condition points, as shown in Fig.~\ref{fig5} and Table~\ref{table3-cost}, the relative optimality and accuracy of IDP and DP methods can be compared. As depicted, in IDP method, the produced optimal policies controlled system with 19 \% less cost value, compared to conventional DP method.
For the IDP algorithm, the first set of ICHS agents can be computed by linearizing the system about the small neighborhood of the origin and solving the resulted LQR problem. Additionally, this approach is applied to the inverted pendulum system as shown in Fig.~\ref{fig5}. The results suggest that the accuracy can be increased by comparing randomly-initialized IDP with LQR and LQR-started IDP. Table~\ref{table3-cost} summarizes the effects of approximating the first \textit{ICHS*} when applying the IDP algorithm.
\begin{table}[t]
	\caption{Parameters of DP and IDP used in system Eq.~(\ref{eq16}) and inverted pendulum (Eq.~(\ref{eq14})) examples.}
	\label{table1-dp param}
	\centering
	\begin{tabular}{|c |c|}
		
		\hline
		\textbf{Parameter} & \textbf{value}   \\

		\hline\hline
		Interval for $x_{1}$ and $x_2$ (DP)  & [-10 10]   \\
		
		Number of grid points for $x_1$ and $x_2$ (DP) & 40    \\
		
		Interval for $u$  (DP and IDP) & [-50 50]   \\
		
		Number of grid points for $u$ (DP) & 30    \\
		
		Decay after one time constant (DP) & 0.9    \\ 
		
		Convergence tolerance (DP and IDP) & $10^{-5}$    \\

		
		Maximum iterations (DP and IDP)  & 10000    \\
			
		Mutation percentage (IDP)  & 0.03   \\
				
		Number of initial agents $N$ (IDP)  &  600   \\
		
		Initial ICHS cost value $\gamma_{0}$ (IDP)  &  0.1   \\
				
		Radius of initial random surface $R_0$ (IDP)  &  0.01   \\
		
		Terminal Cost Value $\gamma_{f}$ (IDP)  &  250  \\
		\hline
		
	\end{tabular}
\end{table}
\begin{figure}[t]
	\begin{center}
		\includegraphics[scale=0.5]{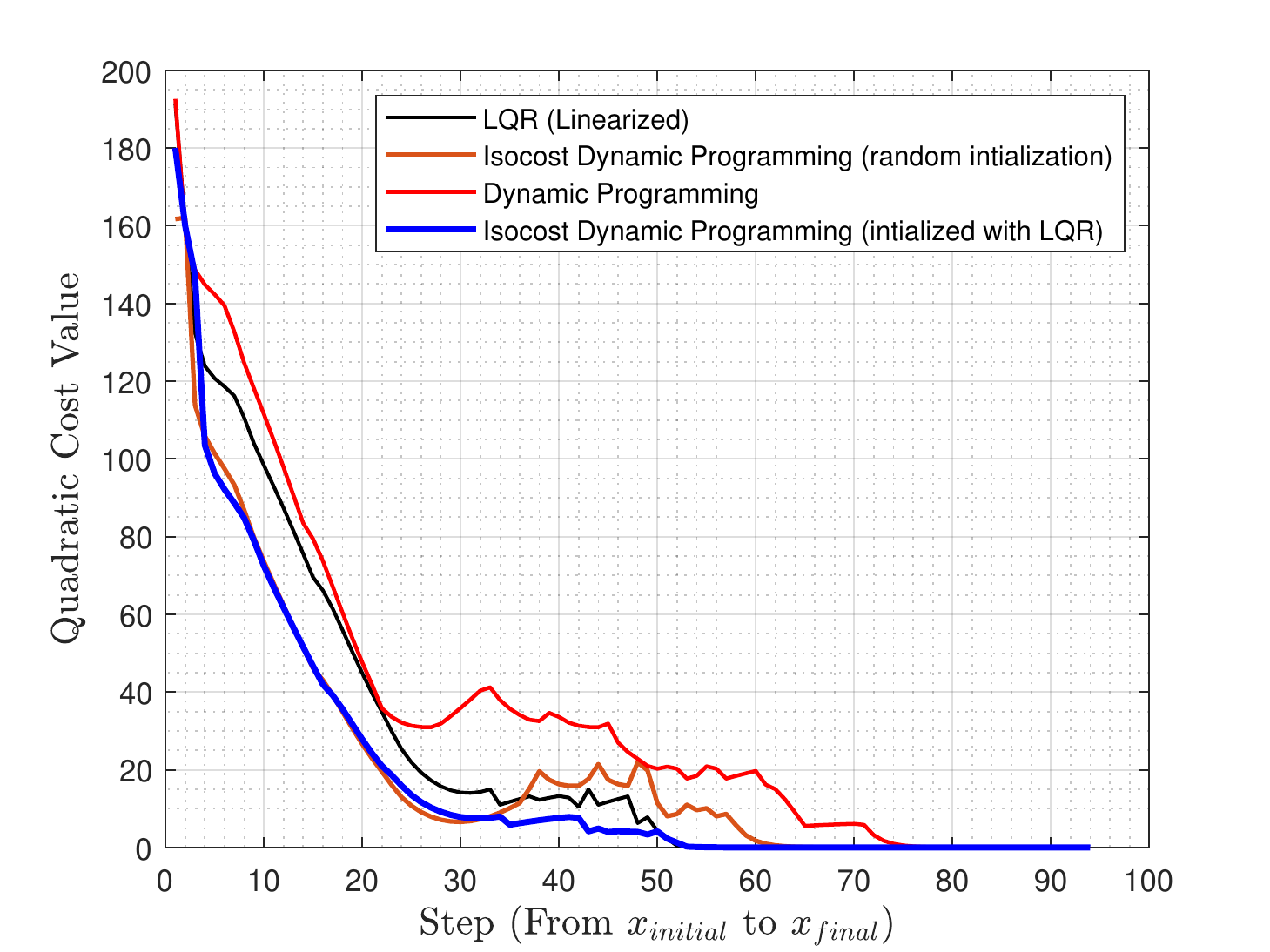}    
		\caption{Changes of cost value in each step for inverted pendulum system controlled with three different control method.}
		\label{fig5} 
	\end{center}                                 
\end{figure}
\begin{figure}[h]
	\begin{center}
		\includegraphics[scale=0.5]{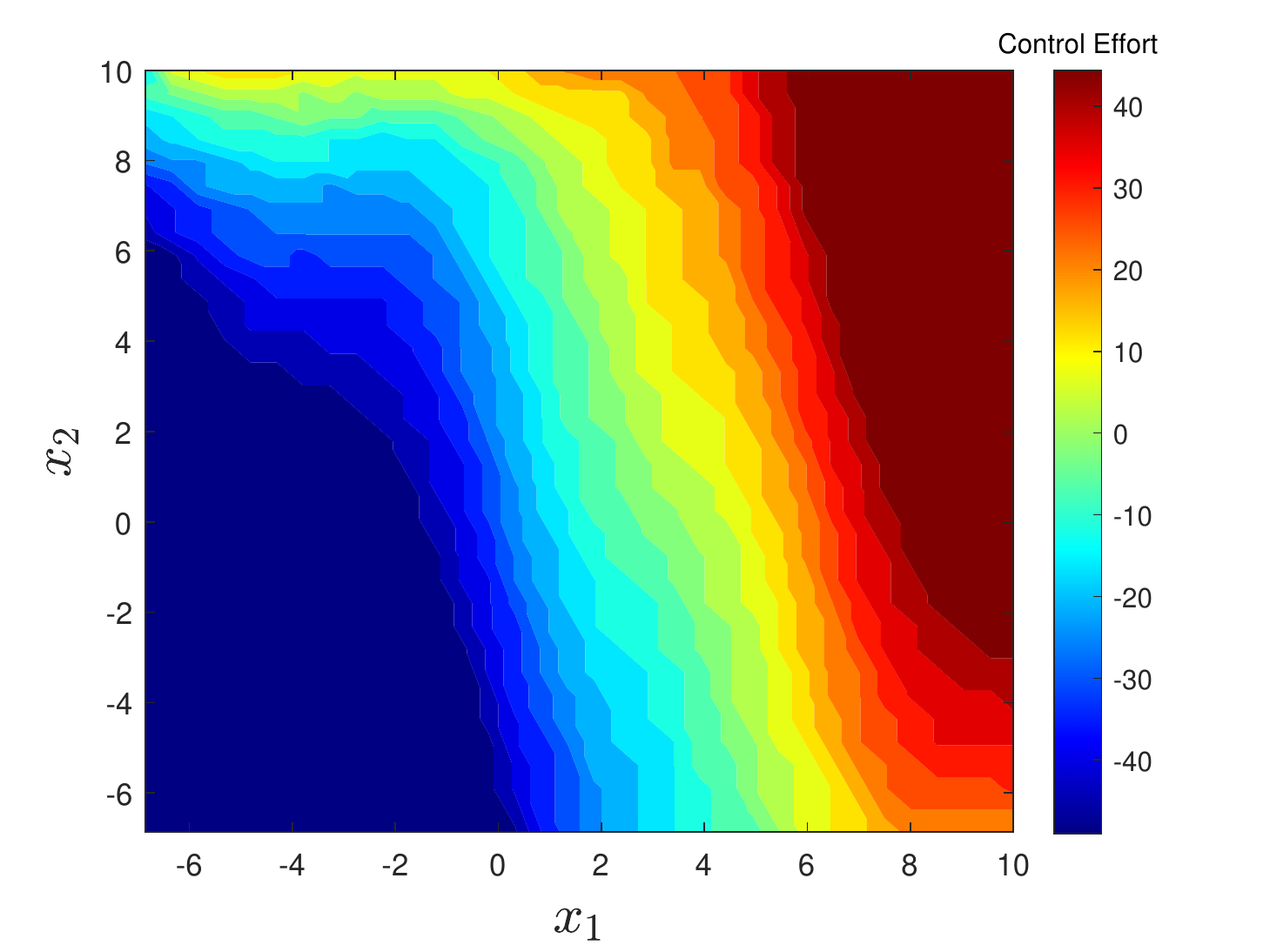}    
		\caption{Optimal control effort value for each state point in a inverted pendulum system controlled with IDP.}  
		\label{fig6}
	\end{center}                                 
\end{figure}
\newline
According to Fig.~\ref{fig6},  once the optimal control value for each agent on the ICHS has been determined, all of these values come together to form a look-up table that aids in determining the optimal control for all other points in the state space of the system.
The IDP controller may therefore respond with the optimal input at every state on the continues space. Then, to find optimal control solution for any points, the 4th order Runge-Kutta method and Barycentric interpolation is used to calculate and save all state transitions~\cite{munos1998barycentric}.  
\newline
To compare convergence time between DP and IDP, which shows the computational load of each method, the inverted pendulum system is controlled with both methods for the same state space intervals and convergence tolerance (as indicated in Table~\ref{table1-dp param}). The result shows that the proposed IDP method perform 21 \% faster, on total time for each run, and 12 \% faster on each iteration, compared to DP, as highlighted in Table~\ref{table2-time}. It worth mentioning that, for the sake of simplicity, the chosen interval for DP was chosen to be small to reduce simulation time. According to Table~\ref{table2-time}, comparing average the time spent one each iteration, it can be concluded that if the intervals for states and number of grid points would increase, the computational speed difference between DP and IDP would increase. Based on the number of grid points for DP (40x30=1200 states) and the number of agents for IDP (600 states), the time difference between these two methods can be explained, as IDP uses half the state points as DP. It also results in using less storage memory. 
\begin{table}[t]
	\caption{Average cost value (average of integral of cost function) of controlled inverted pendulum for 11 random initial conditions.}
	\label{table3-cost}
	\centering
	\begin{tabular}{|c|c|}
		
		\hline	
		\textbf{Control Method} & \textbf{Cost Value}    \\
		
		\hline\hline
		
		IDP ($ S_{\gamma_0}\left(\boldsymbol{x}, c^{*}(\boldsymbol{x})\right)$ from random points) & 2472     \\
		
		IDP ($ S_{\gamma_0}\left(\boldsymbol{x}, c^{*}(\boldsymbol{x})\right)$ from LQR) &  1928 \\
		
		LQR (Linearized System) & 2434    \\
		
		Standard DP & 3028   \\
		\hline
		
	\end{tabular}
\end{table}
\begin{table}[t]
	\caption{Comparing computational performance of DP and IDP control methods for inverted pendulum system.}
	\label{table2-time}
	\centering
	\begin{tabular}{|c |c| c|}
		\hline
	\textbf{Performance Category} & \textbf{DP} & \textbf{IDP}    \\

		\hline\hline
		number of runs & 20  & 20   \\
		
		average time for each run (second) & 840 & 660    \\
		
		average convergence iteration &  1325 &  1187   \\
		
		average time for each iteration (second) & 0.64  &  0.56  \\
		\hline
	\end{tabular}
\end{table}
\section{Conclusion}\label{sec concl}
In this paper, a geometric feature in the state space of a system, called  IsoCost HyperSurface (ICHS)  was introduced. This feature is exploited in the Dynamic Programming (DP) algorithm to solve infinite horizon optimal control problems. It was proved that for a certain amount of cost, the ICHS obtained from an optimal control solution surrounds any other ICHS resulting by a non-optimal control solution.
Then, the resulted theorem and IDP algorithm were applied to two optimal control problems. The results demonstrated that the introduced geometric ICHS feature with an evolutionary-based search method of IDP controller can be used as a capable alternative solver for optimal control problems, increasing the overall accuracy and convergence speed compared to conventional DP.
\newline
Furthermore, the proposed IDP algorithm decreased the computational load (compared to conventional DP), by decreasing the convergence time by 21 \%. To improve the performance of the suggested IDP controller and reduce memory usage, the concept of Deep Neural Networks (DNN) can be incorporated in the IDP controller in future works. Considering the ability of DNNs to serve as function approximators, the DNNs can learn optimal control solutions for the state space of the system from achieved ICHSs for different cost values.
\bibliographystyle{elsarticle-num}        
\bibliography{autosam}                    
\end{document}